\documentclass[smallextended]{svjour3}
\usepackage{graphicx}
\begin{document}
\title{Microwave Optomechanically Induced Transparency and Absorption Between 250 and 450 mK}

\author{Sumit Kumar \and Dylan Cattiaux \and Eddy Collin \and Andrew Fefferman \and Xin Zhou}

\institute{Sumit Kumar \and Dylan Cattiaux \and Eddy Collin \and Andrew Fefferman \at
              Universit\'{e} Grenoble Alpes and Institut N\'{e}el, CNRS, Grenoble, France \\
              \email{andrew.fefferman@neel.cnrs.fr}           %  \\
	\and
	Xin Zhou \at
	Univ. Lille, CNRS, UMR 8520 -
			IEMN, Lille, France
}

\date{\today}

\maketitle

\begin{abstract}
High-quality microwave amplifiers and notch-filters can be made from microwave optomechanical systems in which a mechanical resonator is coupled to a microwave cavity by radiation pressure. These amplifiers and filters rely on optomechanically induced transparency (OMIT) and absorption (OMIA), respectively. Such devices can amplify microwave signals with large, controllable gain, high dynamic range and very low noise. Furthermore, extremely narrowband filters can be constructed with this technique. We briefly review previous measurements of microwave OMIT and OMIA before reporting our own measurements of these phenomena, which cover a larger parameter space than has been explored in previous works. In particular, we vary probe frequency, pump frequency, pumping scheme (red or blue), probe power, pump power and temperature. We find excellent agreement between our measurements and the predictions of input/output theory, thereby guiding further development of microwave devices based on nanomechanics.
\end{abstract}

\section{Introduction}

Superconducting microwave circuits have had a strong impact on science and
technology in recent years. Initial experiments in circuit cavity quantum
electrodynamics \cite{Vion02,Wallraff04} have led to massive efforts in the
field of quantum information processing with superconducting
qubits \cite{Arute19}. Superconducting circuits are also used as highly
sensitive astrophysical detectors \cite{Day03}. At the same time, coupling
mechanical and electromagnetic degrees of freedom via radiation pressure has
led to a plethora of fascinating results. Initially light at (nearly) visible
wavelengths was employed for these optomechanical
studies \cite{Cohadon99,Carmon05,Kleckner06,Arcizet06,Poggio07}.

More recently, the techniques of superconducting microwave circuits and
optomechanics were combined by shifting from visible light to microwave
frequencies \cite{Regal08}. This was followed by impressive achievements in
areas including quantum control of mechanical resonators \cite{OConnell10} as
well as microwave circulators \cite{Bernier17,Barzanjeh17} and
amplifiers \cite{Massel11}.

Microwave amplifiers and notch filters can rely on optomechanically induced
transparency (OMIT) and absorption (OMIA), respectively. These phenomena are
observed when a strong pump and a weak probe with a frequency difference near
the mechanical resonance frequency are simultaneously applied to the
electromagnetic resonator, resulting in a change in the transmission of the
probe. Several workers have reported measurements of OMIT and OMIA. Weis
\textit{et al}. \cite{Weis10} observed OMIT in toroidal
whispering-gallery-mode microresonators illuminated by a laser operating at a
wavelength of 775 nm. In this work the dependence of the probe transmission on
pump power and detuning was investigated. Safavi-Naeini \textit{et al}.
\cite{Safavi11} observed OMIT and OMIA in optomechanical crystals illuminated
by laser light with wavelength of 1550~nm. Here the pump power and detuning
was varied, and measurements were made at 8.7 kelvin and at room temperature.

In one of the first demonstrations of OMIT at microwave frequencies, Teufel
\textit{et al}. \cite{Teufel11a} presented highly sensitive measurements of
the probe transmission upon varying the pump power and detuning. Shortly thereafter, an analysis
of the use of OMIT for microwave amplification, as well as the observation of
OMIA at microwave frequencies, was presented by Massel \textit{et al.} \cite{Massel11}. In that work, the dependence of the probe transmission on
pump power was studied. Hocke \textit{et al.} focused on microwave OMIA and reported
transmission as a function of detuning and drive power up to the mechanical
parametric instability \cite{Hocke12}. Zhou \textit{et al.} returned to microwave OMIT
and scanned the probe power, thereby driving the mechanical resonator strongly
enough so that its Duffing non-linearity became significant \cite{Zhou13}. The
ability of microwave optomechanical amplifiers to function at high probe power
was later confirmed with a relatively elaborate device \cite{Ockeloen16}.
Recent work has demonstrated the advantage of combining mechanical parametric
amplification with OMIT \cite{Bothner20}\ as well as microwave amplification
in the absence of the dynamical backaction responsible for amplification of
mechanical motion \cite{Cohen20}. The lack of dynamical backaction has the advantage that quantum-limited microwave amplification could be achieved even in the presence of large thermal occupation of the mechanical mode \cite{Cohen20}.

Here we report the results of our two tone measurements of a microwave
optomechanical device, in which we explore a larger parameter space than has
been covered in previous works. In particular, we vary probe frequency, pump
frequency, pumping scheme (red or blue), probe power, pump power and
temperature. We demonstrate that the theoretical transmission derived from
input/output theory is in excellent agreement with our measurements over the
entire parameter range.

\section{Optomechanically induced transparency and
absorption\label{sec:theory}}

\begin{figure}
\includegraphics[width=\textwidth]{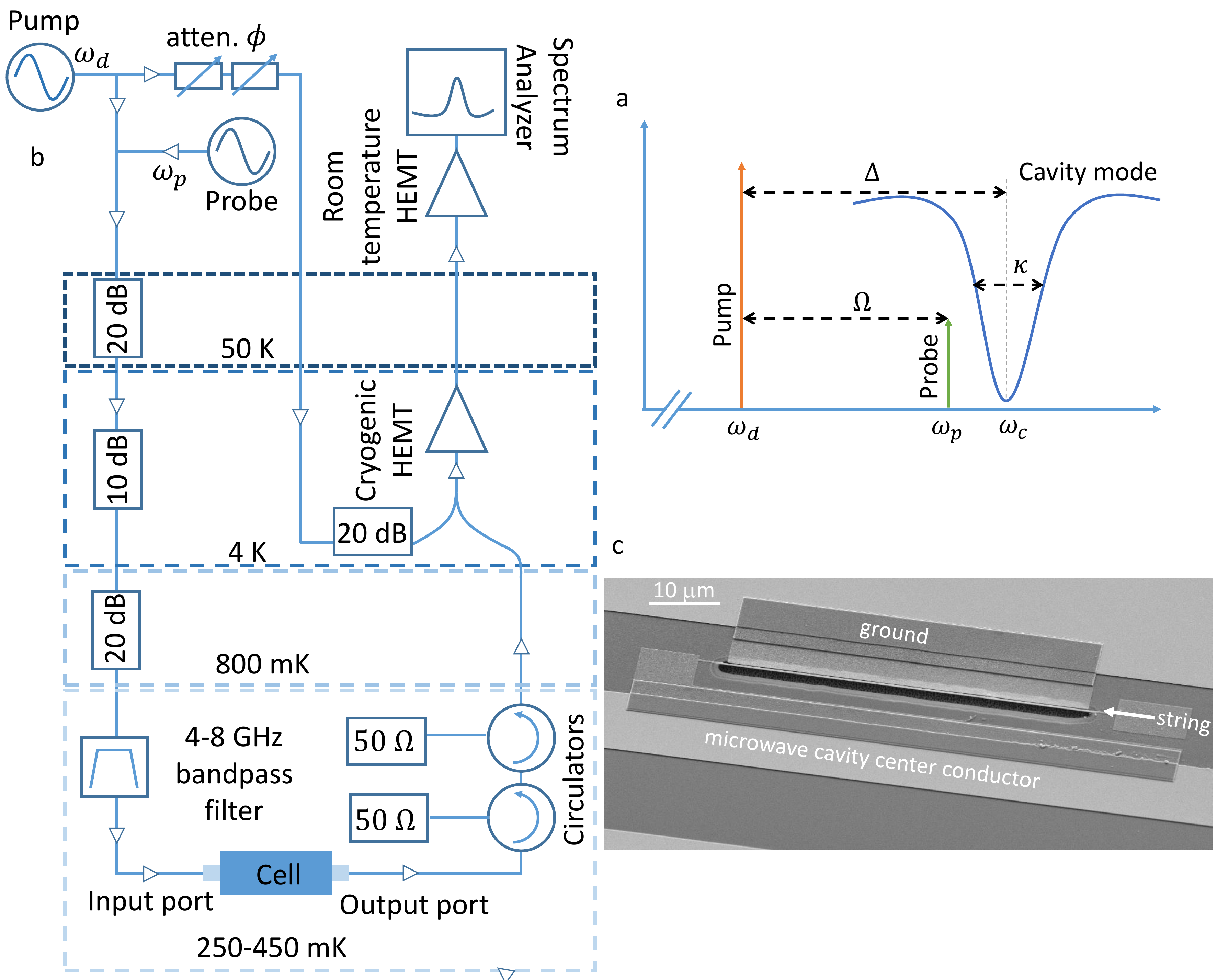}
\caption{(a) The pump and probe scheme, (b) the microwave circuit and (c) the nanomechanical resonator, which is coupled to the microwave cavity.}
\label{fig:setup}
\end{figure}

OMIT and OMIA can be observed in systems where a mechanical resonator is
coupled to an electromagnetic resonator (the cavity) in such a way that the
motion of the former changes the resonance frequency of the latter. The
interaction between a single phonon and a single photon is characterized by
$g_{0}$, the vacuum optomechanical coupling strength, which is the frequency
shift of the cavity due to zero point motion of the mechanical resonator. When
the cavity is driven by a pump at frequency $\omega_{d}$ and a probe at
frequency $\omega_{p}$ (Fig. \ref{fig:setup}a), there is a component of the resulting
radiation pressure force at the difference frequency $\Omega=\omega_{p}%
-\omega_{d}$. We denote the mechanical resonance frequency by $\Omega_{m}$.
When $||\Omega|-\Omega_{m}|$ is less than or approximately equal to the
mechanical linewidth, the response of the mechanical element is appreciable.
Now its motion yields up-conversion or down-conversion of the photons by an
amount $\left\vert \Omega\right\vert $, yielding sidebands in the microwave
spectrum. One of the two sidebands of the pump is at a frequency that
coincides with the probe frequency. Whether this is the upper or lower
sideband depends on the sign of $\Omega$. The probe and the coincident sideband
interfere constructively or destructively depending on their relative phases,
resulting in a change in the transmission of the probe relative to the value
it would have in the absence of optomechanical coupling. This effect is most
significant when the probe frequency is within about one half linewidth
$\kappa$ of the cavity resonance frequency $\omega_{c}$. This is equivalent to
$|\Delta+\Omega|<\approx\kappa/2$, where we define $\Delta=\omega_{d}%
-\omega_{c}$.

The magnitude of the response of the mechanical resonator to the radiation
pressure at the beat frequency $\Omega$ influences the probe transmission. The
mechanical response depends not only on the difference between the driving
frequency and the mechanical resonance $\left\vert \Omega\right\vert
-\Omega_{m}$ and the intrinsic linewidth of the mechanical resonator
$\Gamma_{m}$ but also on dynamical backaction due to the pump
\cite{Aspelmeyer14}. The backaction yields an effective mechanical linewidth
$\Gamma_{eff}$, which is greater than $\Gamma_{m}$ for $\omega_{c}-\omega
_{d}\approx\Omega_{m}$ (\textquotedblleft red pumping\textquotedblright)
and less than $\Gamma_{m}$ for $\omega_{d}-\omega_{c}\approx\Omega_{m}$
(\textquotedblleft blue pumping\textquotedblright).

The theoretical probe transmission $S_{21}$ in the presence of a pump,
consistent with the qualitative discussion of OMIT and OMIA above, has been
derived in previous works \cite{Agarwal10,Zhou13}. The output from the
microwave cavity $\hat{a}_{\mathrm{out}}$ is related to the applied field $%
\hat{a}_{\mathrm{in}}$ by \cite{Gardiner04}%
\[
\hat{a}_{\mathrm{out}}=\hat{a}_{\mathrm{in}}+\sqrt{\kappa _{ext}/2}\ \hat{a}
\]%
where $\kappa _{ext}$ is the part of the cavity linewidth due to coupling to
the feedline and $\hat{a}$ is the cavity field. This input/output relation,
along with the equations of motion for the microwave cavity field \cite%
{Clerk10} and the mechanical oscillator position $x$, is the starting point
for the derivation of the probe $S_{21}$\cite{Zhou13}. We define the respective cavity and mechanical susceptibilities as $\chi_{c}^{-1}\left(  \omega\right) =\kappa/2-i\left(\omega-\omega_{c}\right)$ and $\chi_{m}^{-1}\left(  \omega\right)  =\Gamma_{m}/2-i\left(\omega-\omega_{d}\pm\Omega_{m}\right)$ where the upper (lower) sign is for blue (red) pumping. Then
\begin{equation}
S_{21}\left(  \omega_{p}\right)  =1-\frac{\chi_{c}\left(  \omega_{p}\right)  \kappa_{ext}/2}{1\mp
g_{0}^{2}~n_{cav}\chi_{c}\left(  \omega_{p}\right)  \chi_{m}\left(  \omega
_{p}\right)  }\label{transmission}
\end{equation}
where
\begin{eqnarray}
\chi_{c}^{-1}\left(  \omega_{p}\right)   & =\frac{\kappa}{2}-i\left(\Omega+\Delta\right), \nonumber\\
\chi_{m}^{-1}\left(  \omega_{p}\right)   & =\frac{\Gamma_{m}}{2}-i\left(\Omega\pm\Omega_{m}\right),\nonumber
\end{eqnarray}
and
\begin{equation}
n_{cav}  =\frac{P_{in}\kappa_{ext}}{2\hbar\omega_{d}}|\chi_{c}\left(\omega_{d}\right)  |^{2}.
\end{equation}
Here $\kappa_{ext}$ is the
part of the cavity linewidth due to intentional coupling to the feedline,
$n_{cav}$ is the number of photons stored in the cavity due to the pump, and
$P_{in}$ is the pump power at the input of the optomechanical device.

\section{Experiment}

Our optomechanical device and microwave measurement circuit are similar to
ones used in previous works, e.g., \cite{Regal08,Zhou19}. A circuit diagram is
shown in Fig. \ref{fig:setup}b. The pump and probe tones are combined at room
temperature. These signals then pass through attenuators, which decrease the
thermal noise level, and a bandpass filter before reaching the
input port of the experimental cell. The bandpass filter is used to remove spurious signals coming from, e.g., the microwave generators. The transmitted signal that exits the
cell passes through two circulators, which prevent noise traveling down the
detection line from entering the output port of the cell. At the 4 kelvin
plate, the pump signal that was transmitted through the cell is canceled to
avoid saturating the amplifiers. The resulting signal passes through a
cryogenic high electron mobility transistor (HEMT) amplifier and a room
temperature HEMT amplifier before measurement with a spectrum analyzer.

The experimental cell is a box made of annealed CuC2. It is pressed against
the mixing chamber plate of the cryogen-free dilution refrigerator. Inside the
cell the input and output coaxial transmission lines are soldered to gold
coplanar waveguides (CPW) on a circuit board, which are in turn microbonded to
the ends of a niobium CPW on a chip. The chip is made of silicon and is coated
with 100 nm of high stress silicon nitride. A $\lambda/4$ CPW resonator is
also present on the chip and forms a cavity with resonance frequency
$\omega_{c}/2\pi=6$ GHz. The open end of the $\lambda/4$ resonator is
precisely situated relative to the CPW feedline to achieve the desired
coupling. The strength of the coupling is characterized by an external cavity
linewidth $\kappa_{ext}/2\pi=44$ kHz. The total cavity linewidth is
$\kappa/2\pi\approx100$ kHz, so that the condition for critical coupling $\kappa=2\kappa_{ext}$ is nearly satisfied.

The mechanical element is a vibrating string made from silicon nitride at the
open end of the cavity (Fig. \ref{fig:setup}c). Its geometry was defined by a
30 nm thick aluminum layer that served as a reactive ion etch (RIE) mask.
After the anisotropic RIE, the string was released by selective XeF$_{2}$
etching of the silicon substrate. The aluminum was not removed after etching,
and it yields electrostatic coupling between the vibrating string and the
cavity characterized by the coupling strength $g_{0}/2\pi=0.56$ Hz. The string
has a length of 50 $\mu$m and a resonance frequency $\Omega_{m}/2\pi=3.8$ MHz.

\section{Results and Discussion}

\begin{figure}
\includegraphics[width=\textwidth]{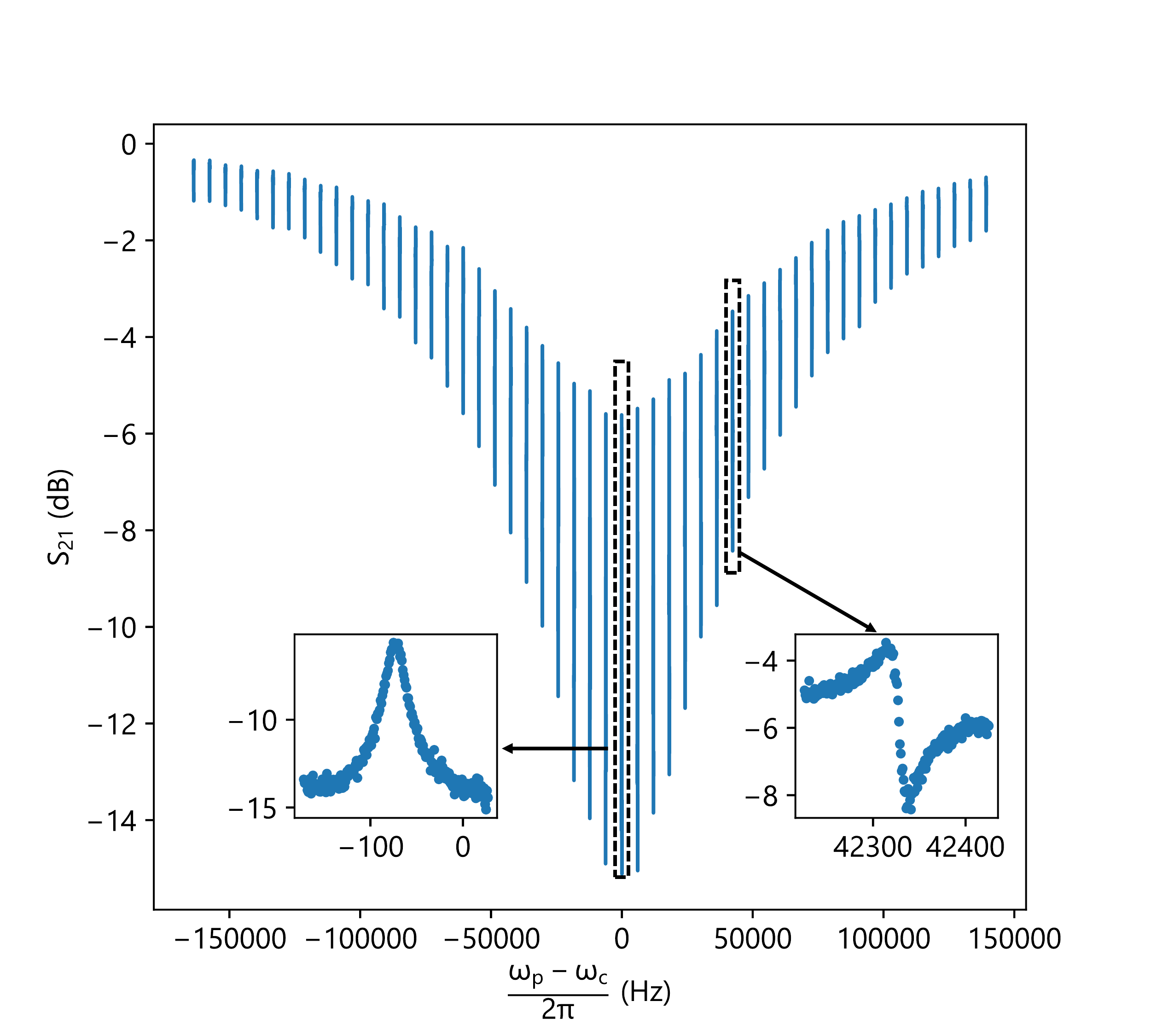}
\caption{Probe transmission measurements at 250 mK with red pumping
yielding $n_{cav}=1.3\times10^{6}$ and a probe power of -116 dBm at the input
of the cell. The narrow probe frequency sweeps appear as vertical
lines. The insets show enlargements of the indicated sweeps.}
\label{fig:measurement}
\end{figure}

The measurements were carried out at sample temperatures of 250, 350 and 450
mK. For each temperature, pump power and probe power setting, the pump and
probe frequencies were scanned and the probe transmission was measured. Figure
\ref{fig:measurement} shows our measurements at 250 mK with red pumping
yielding $n_{cav}=1.3\times10^{6}$ and a probe power of -116 dBm at the input
of the cell. The transmission measurements were made by setting $\omega_{d}$
and sweeping the probe frequency $\omega_{p}$ over a narrow range around
$\omega_{d}+\Omega_{m}$ for red pumping and around $\omega_{d}-\Omega_{m}$ for
blue pumping. The width of the probe frequency sweeps was comparable to the
mechanical linewidth, so that it encompassed the OMIT/OMIA resonance. These
measurements were made at a range of pump frequencies $\omega_{d}$ such that
the set of center frequencies of the narrow probe frequency sweeps spanned the
microwave resonance. These narrow sweeps appear as vertical lines on the scale
of the main panel of Fig. \ref{fig:measurement}; two of the sweeps are
enlarged in the insets.

\begin{figure}
\includegraphics[width=\textwidth]{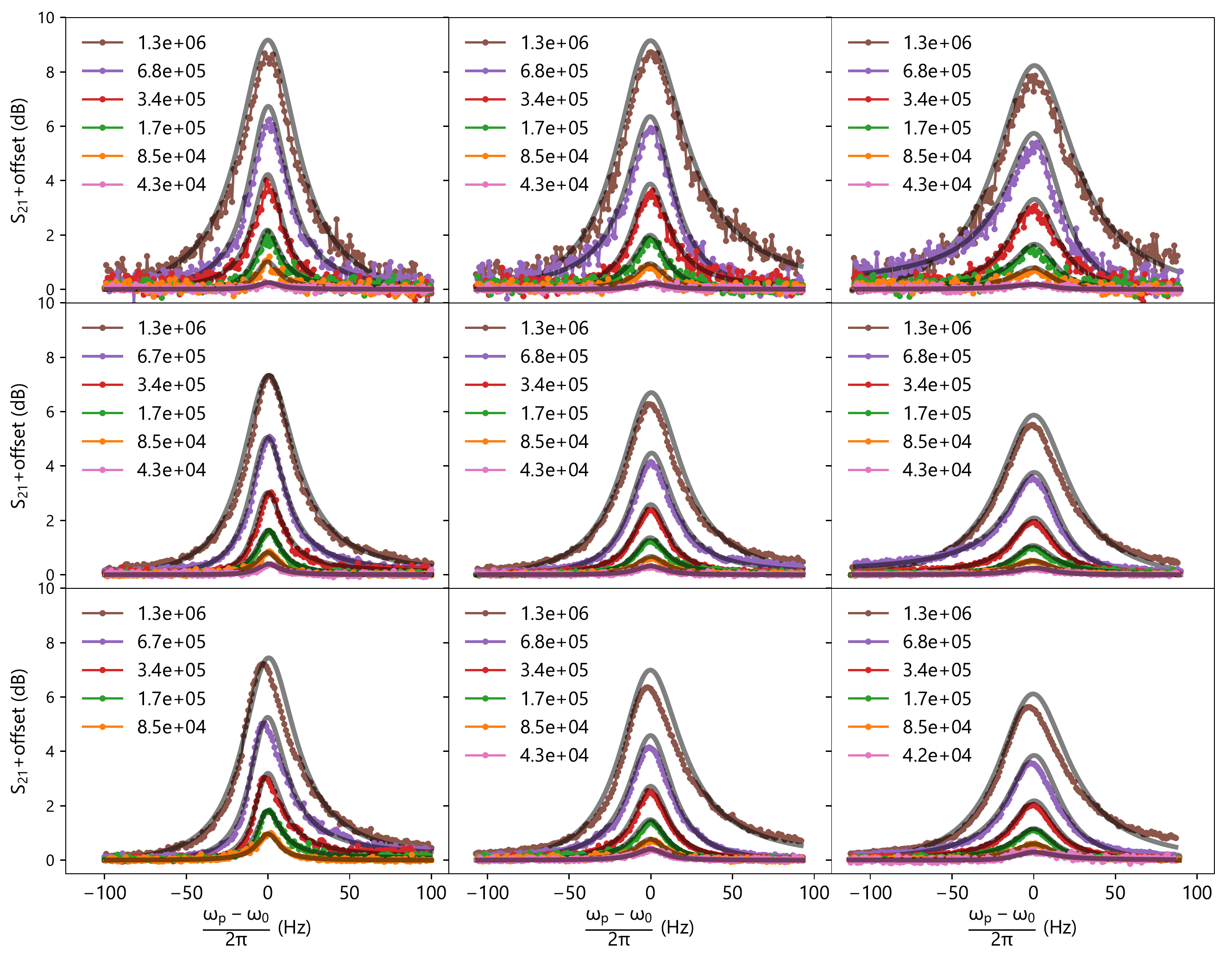}
\caption{Main panel: Probe transmission measurements at the indicated
temperatures and probe powers referenced to the input of the cavity. Red
pumping was applied, yielding the specified intracavity pump photon numbers.
The transmission curves are offset vertically for clarity and $\omega_0$ is an offset close to $\omega_c$. Grey curves are fits of Eq. \ref{transmission} to the data.}
\label{fig:redpumpcuts}
\end{figure}

Figure \ref{fig:redpumpcuts} shows OMIT\ at the microwave resonance. Any
residual detuning is very small compared with the cavity linewidth. These data
were acquired under red pumping ($\omega_{d}=\omega_{c}-\Omega_{m}$), and the
dependence on temperature, probe power and pump power is shown. The intracavity
pump photon numbers and probe powers at the input of the cavity are based on a
careful measurement of the attenuation of the transmission line connecting the
generator to the microwave cavity. At constant temperature and probe power the
amplitude of the OMIT resonance increases with pump power because (1) the
mechanical mode is driven more strongly and (2) more pump photons are
available for up-conversion to the probe frequency by interaction with the
mechanical mode. The resulting upper sideband interferes constructively with
the probe. The width of the OMIT resonance increases with pump power due to
dynamical backaction on the mechanical mode. The curves represent fits of the
theoretical $|S_{21}|$ (Eq. \ref{transmission}) to the data. Single values of
$\Omega_{m}$ and $\Gamma_{m}$ were chosen for each temperature to optimize the
fits of the theoretical transmission to the entirety of our measurements of
this device. At 350 (450) mK, $\Omega_{m}/2\pi$ is higher than its 250 mK
value by 7 (12) Hz. The intrinsic mechanical linewidths $\Gamma_{m}/2\pi$ at
250, 350 and 450 mK are 15.3, 20.0 and 26.8 Hz, respectively. The values of
$\omega_{c}$ and $\kappa$ were allowed to vary to account for the dependence
of these parameters on cavity temperature and photon population. The
dependence of $\kappa$ on probe power is primarily responsible for the
dependence of the transmission on probe power shown in Fig.
\ref{fig:redpumpcuts}.

\begin{figure}
\includegraphics[width=\textwidth]{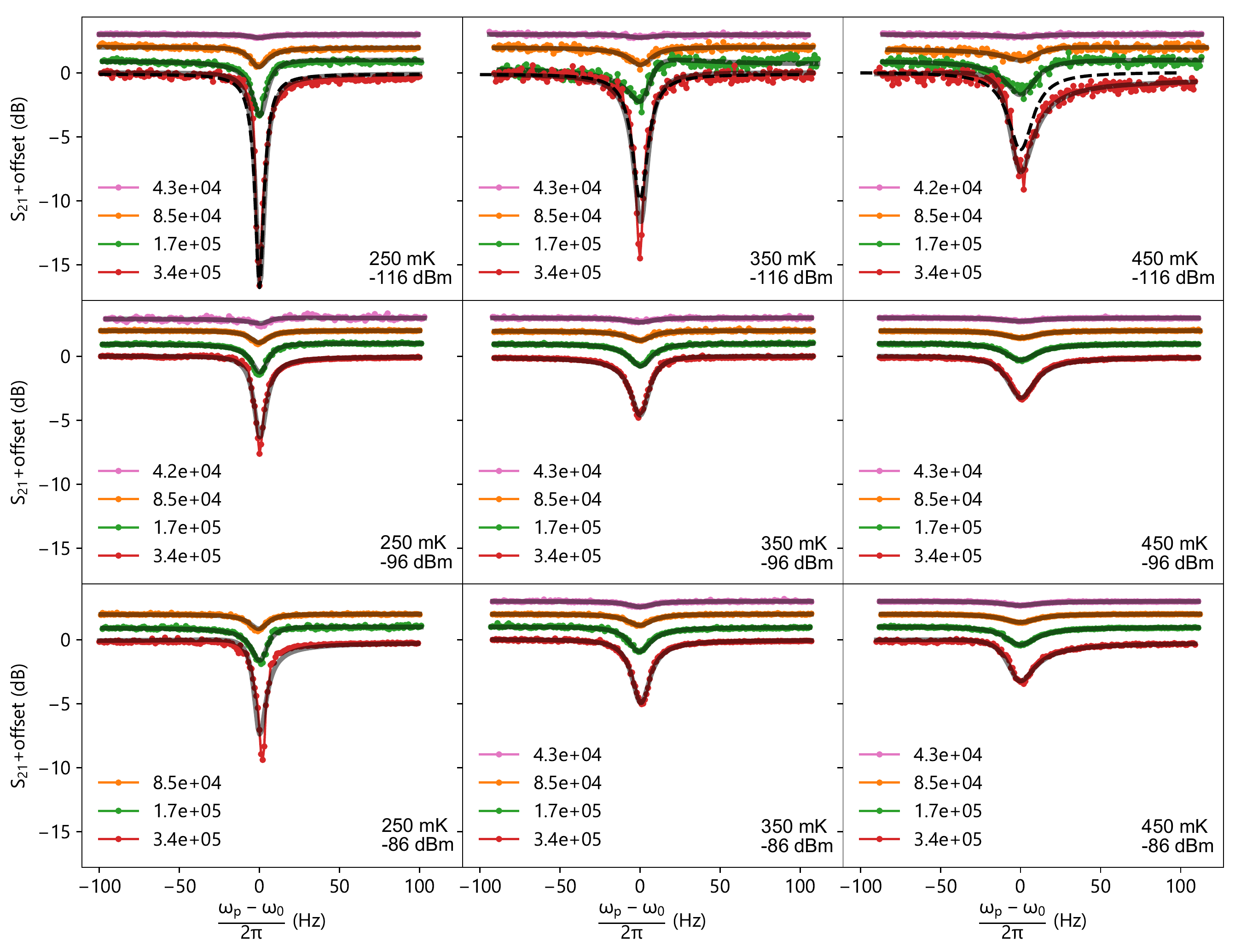}
\caption{Probe transmission measurements at the indicated
temperatures and probe powers referenced to the input of the cavity. Blue
pumping was applied, yielding the specified intracavity pump photon numbers.
The transmission curves are offset vertically for clarity and $\omega_0$ is an offset close to $\omega_c$. The curves are fits of Eq. \ref{transmission} to the data. The dashed black curves show the part of the temperature dependence due to variation of $\Omega_{m}$ and $\Gamma_{m}$ (see text).}
\label{fig:bluepumpcuts}
\end{figure}

Figure \ref{fig:bluepumpcuts} shows OMIA on the microwave resonance at the
same probe powers and temperatures as in Fig. \ref{fig:redpumpcuts}. These
data were acquired under blue pumping ($\omega_{d}=\omega_{c}+\Omega_{m}$). As
for red pumping, the driving force acting on the string and the number of pump
photons stored in the cavity increase with pump power. But for blue pumping
the lower sideband interferes destructively with the probe, yielding decreased
transmission of the probe in Fig. \ref{fig:bluepumpcuts}. Furthermore, the
width of the optomechanical resonance decreases with pump power due to
dynamical backaction on the mechanical mode. At pump powers that are higher
than those shown in Fig. \ref{fig:bluepumpcuts}, $\Gamma_{eff}$ vanishes and
the mechanical mode undergoes self-sustained oscillations. As we approach the auto-oscillation threshold, we observe
a response that is inconsistent with the theory presented here and goes beyond the scope of the present work.

The solid gray curves
represent fits of the theoretical $|S_{21}|$ (Eq. \ref{transmission}) to the
data, where the values of $\Omega_{m}$ and $\Gamma_{m}$ at each temperature
are the same as the ones used to fit the red pumping data (Fig.
\ref{fig:redpumpcuts}). The temperature dependence of $\Omega_{m}$ and $\Gamma_{m}$ is not entirely responsible for the temperature dependence of these data: The dashed black curves in the top row of Fig. \ref{fig:bluepumpcuts} were generated by retaining the values of $\Omega_{m}$ and $\Gamma_{m}$ corresponding to the temperature but setting $\omega_{c}$ and $\kappa$ to their best fit values at 250 mK and maximum pump power. The dependence of $\kappa$ on temperature is responsible for much of the discrepancy between the solid gray and dashed black curves but doesn't account for the small asymmetry about the minimum of the solid gray curve. The latter is due to a small detuning from the cavity resonance.

\begin{figure}
\includegraphics[width=\textwidth]{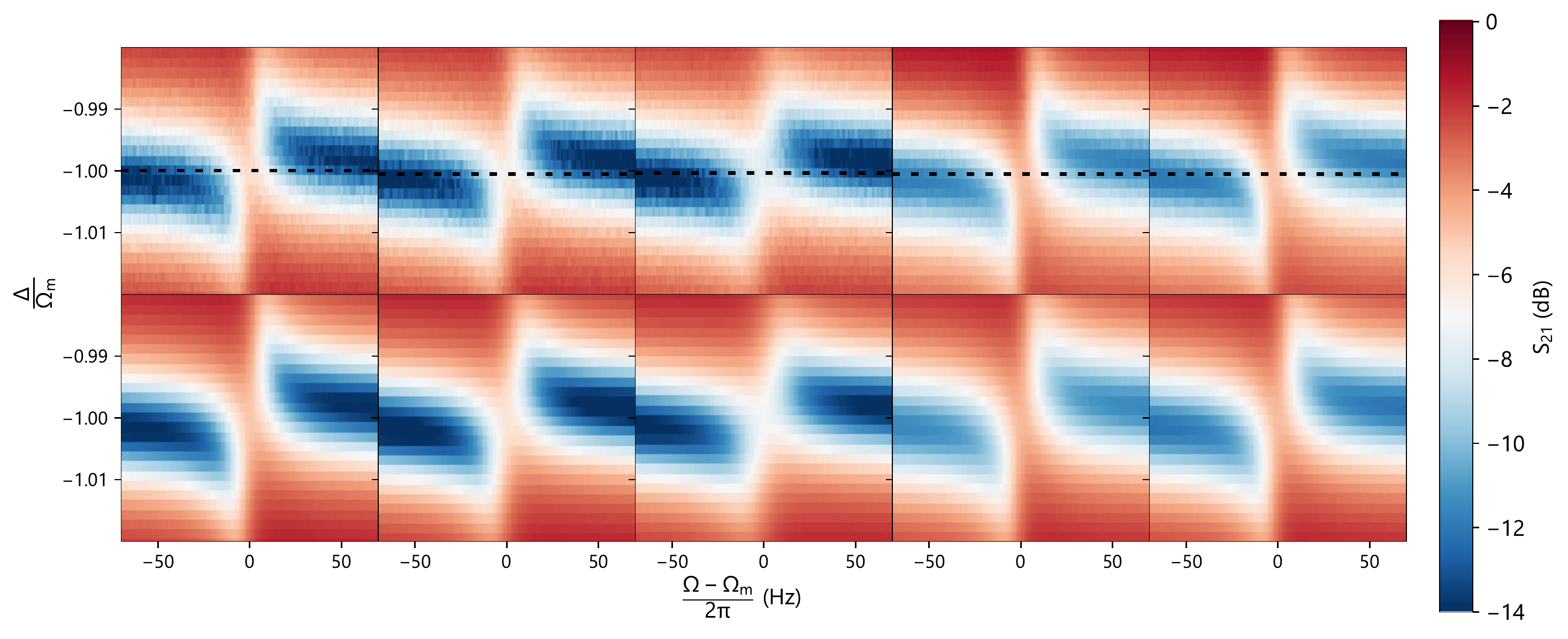}
\caption{Measured (upper panel) and theoretical (lower panel) probe transmission at the maximum red pumping power ($n_{cav}=1.3\times10^{6}$). From left to right the values of [temperature (mK), probe power (dBm),$\kappa/2\pi$ (kHz) and $\Delta\omega_c/2\pi$ (kHz)] are [250,-116,84,0],[350,-116,82,52],[450,-116,83,93],[250,-96,96,-10],[250,-86,95,-9], where $\Delta\omega_c$ is the shift in $\omega_c$ relative to its value in the leftmost panels.  The black dashed lines correspond to data shown in the top row and left column of panels of Fig. \ref{fig:redpumpcuts}.}
\label{fig:redmap}
\end{figure}

The theoretical transmission is also in excellent agreement with our
measurements made further from the microwave resonance. The top row of panels
in Fig. \ref{fig:redmap} shows $|S_{21}|$ as a function of $\Omega$ and
$\Delta$ at the maximum red pumping power ($n_{cav}=1.3\times10^{6}$) and the
specified temperatures and probe powers. Horizontal line cuts in this figure correspond to
particular pump frequencies $\omega_{d}$ and narrow sweeps of $\omega_{p}$ centered on
$\omega_{d}+\Omega_{m}$. Thus each spectrum appearing as a vertical line in Fig. \ref{fig:measurement} corresponds to a horizontal line-cut at a particular value of
$\Delta$ in the upper-left panel of Fig. \ref{fig:redmap}.

In Figs. \ref{fig:measurement} and \ref{fig:redmap}, the OMIT signal decreases
significantly, as expected, when the frequency of the upper mechanical
sideband of the pump is not well-aligned with the microwave resonance. In Fig.
\ref{fig:measurement} this condition corresponds to the left and right
extremities of the plot and in Fig. \ref{fig:redmap} it corresponds to the
upper and lower extremities of each panel. The fact that the size of the OMIT resonance in Fig. \ref{fig:redmap} is largest for $|\Delta/\Omega_{m}+1|<\approx 0.5\kappa/\Omega_{m}\approx10^{-2}$ follows from the condition $|\Delta+\Omega|<\approx\kappa/2$ (Section \ref{sec:theory}).

The black dashed lines in Fig. \ref{fig:redmap} are at the optimal pump
detuning $\Delta=-\Omega_{m}$ and correspond to data shown in the top row and
left column of panels of Fig. \ref{fig:redpumpcuts}. Note that for the pump
detuning $\omega_{d}=\omega_{c}-\Omega_{m}$ used in Fig. \ref{fig:redpumpcuts}
the quantity on the x-axis of that figure ($\omega_{p}-\omega_{c}$) is
equivalent to the one on the x-axis of Fig. \ref{fig:redmap} ($\Omega
-\Omega_{m}$). Thus we have already demonstrated excellent agreement between
theory and the data along the black dashed lines in Fig. \ref{fig:redmap}. The
panels in the bottom row of Fig. \ref{fig:redmap} demonstrate the same level
of agreement for the case where the upper mechanical sideband of the pump is
slightly detuned from the microwave resonance. These panels show $|S_{21}|$
from Eq. \ref{transmission} with the values of $\Omega_{m}$ and $\Gamma_{m}$
corresponding to the temperatures given in the respective upper panels
and the indicated best-fit values of $\kappa$ and $\omega_{c}$. The increase in $\kappa$ from 84 kHz to 96 kHz as the probe power is increased from -116 dBm to -96 dBm is comparable to the small fluctuations observed in the dependence of microwave cavity loss on drive power observed in Ref. \cite{OConnell08}.

\begin{figure}
\includegraphics[width=\textwidth]{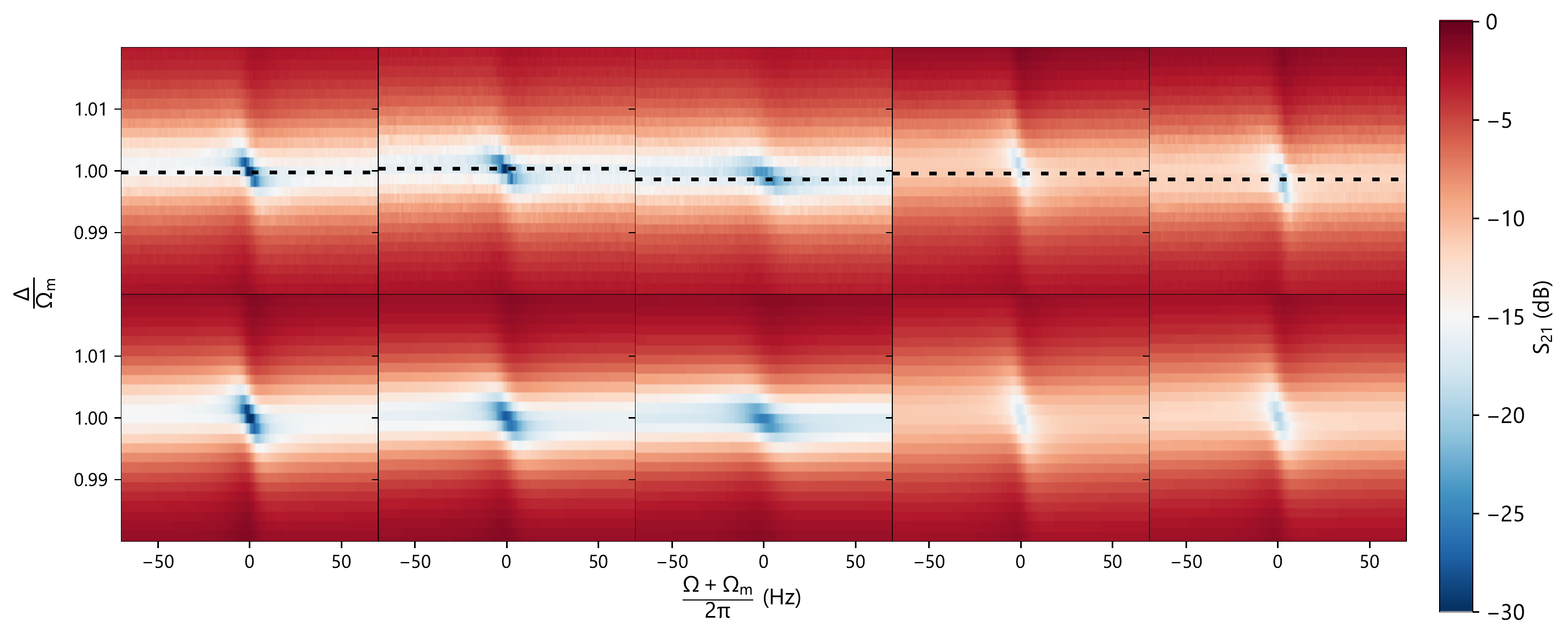}
\caption{Measured (upper panel) and theoretical (lower panel) probe transmission at the maximum blue pumping power ($n_{cav}=3.4\times10^{5}$). From left to right the values of [temperature (mK), probe power (dBm),$\kappa/2\pi$ (kHz) and $\Delta\omega_c/2\pi$ (kHz)] are [250,-116,83,0],[350,-116,80,37],[450,-116,78,80],[250,-96,103,-17],[250,-86,98,-19], where $\Delta\omega_c$ is the shift in $\omega_c$ relative to its value in the leftmost panels. The black dashed lines correspond to data shown in the top row and left column of panels of Fig. \ref{fig:bluepumpcuts}.}
\label{fig:bluemap}
\end{figure}

The top row of panels in Fig. \ref{fig:bluemap} shows $|S_{21}|$ as a function
of $\Omega$ and $\Delta$ at the maximum blue pumping power ($n_{cav}%
=3.4\times10^{5}$) and the specified temperatures and probe powers. The panels in the bottom row of Fig.
\ref{fig:bluemap} show $|S_{21}|$ from Eq. \ref{transmission} with the values
of $\Omega_{m}$ and $\Gamma_{m}$ corresponding to the temperatures given in
the respective upper panels and the indicated best-fit values of $\kappa$ and $\omega_{c}$. The agreement between theory and
experiment is again excellent.

\section{Conclusion}

Nanomechanical resonators coupled to microwave cavities by radiation pressure
constitute ideal systems for microwave amplification and absorption. The
amplifiers or filters can have a very low noise level or a very narrow
bandwidth, and the gain or attenuation depends very strongly on pump power.
Our measurements confirm the applicability of the theoretical transmission Eq.
\ref{transmission} to a larger parameter space than has been covered in
previous works. We demonstrated excellent agreement with theory over a wide
range of probe frequencies, pump frequencies, probe powers, pump powers and
temperatures for both red and blue pumping, thereby facilitating further
development of microwave devices based on nanomechanics.

The data used here is available at\\
https://cloud.neel.cnrs.fr/index.php/s/CnnYPKn8XHYZgXa.

\section{Acknowledgments} This work was supported by the ERC StG grant UNIGLASS
No.714692, STaRS-MOC Project No. 181386 from Region Hauts-de-France, ISITE-MOST project No. 201050, and the ERC CoG grant ULT-NEMS No. 647917. The research leading to these results has received funding from the European Union's Horizon 2020 Research and Innovation Programme, under Grant Agreement no 824109.
\\
\providecommand{\newblock}{}


\begin{thebibliography}{10}
\expandafter\ifx\csname url\endcsname\relax
  \def\url#1{{\tt #1}}\fi
\expandafter\ifx\csname urlprefix\endcsname\relax\def\urlprefix{URL }\fi
\providecommand{\eprint}[2][]{\url{#2}}

\bibitem{Vion02}
Vion D, Aassime A, Cottet A, Joyez P, Pothier H, Urbina C, Esteve D and Devoret
  M~H 2002 {\em Science\/} {\bf 296} 886--889

\bibitem{Wallraff04}
Wallraff A, Schuster D~I, Blais A, Frunzio L, Huang R~S, Majer J, Kumar S,
  Girvin S~M and Schoelkopf R~J 2004 {\em Nature\/} {\bf 431} 162--167

\bibitem{Arute19}
Arute F, Arya K, Babbush R, Bacon D, Bardin J~C, Barends R, Biswas R, Boixo S,
  Brandao F~G, Buell D~A {\em et~al.\/} 2019 {\em Nature\/} {\bf 574} 505--510

\bibitem{Day03}
Day P~K, LeDuc H~G, Mazin B~A, Vayonakis A and Zmuidzinas J 2003 {\em Nature\/}
  {\bf 425} 817--821

\bibitem{Cohadon99}
Cohadon P~F, Heidmann A and Pinard M 1999 {\em Phys. Rev. Lett.\/} {\bf 83}(16)
  3174--3177
  \urlprefix\url{https://link.aps.org/doi/10.1103/PhysRevLett.83.3174}

\bibitem{Carmon05}
Carmon T, Rokhsari H, Yang L, Kippenberg T~J and Vahala K~J 2005 {\em Phys.
  Rev. Lett.\/} {\bf 94}(22) 223902
  \urlprefix\url{https://link.aps.org/doi/10.1103/PhysRevLett.94.223902}

\bibitem{Kleckner06}
Kleckner D and Bouwmeester D 2006 {\em Nature\/} {\bf 444} 75--78

\bibitem{Arcizet06}
Arcizet O, Cohadon P~F, Briant T, Pinard M and Heidmann A 2006 {\em Nature\/}
  {\bf 444} 71--74

\bibitem{Poggio07}
Poggio M, Degen C, Mamin H and Rugar D 2007 {\em Physical Review Letters\/}
  {\bf 99} 017201

\bibitem{Regal08}
Regal C, Teufel J and Lehnert K 2008 {\em Nature Physics\/} {\bf 4} 555--560

\bibitem{OConnell10}
O’Connell A~D, Hofheinz M, Ansmann M, Bialczak R~C, Lenander M, Lucero E,
  Neeley M, Sank D, Wang H, Weides M {\em et~al.\/} 2010 {\em Nature\/} {\bf
  464} 697--703

\bibitem{Bernier17}
Bernier N~R, Toth L~D, Koottandavida A, Ioannou M~A, Malz D, Nunnenkamp A,
  Feofanov A and Kippenberg T 2017 {\em Nature communications\/} {\bf 8} 1--8

\bibitem{Barzanjeh17}
Barzanjeh S, Wulf M, Peruzzo M, Kalaee M, Dieterle P, Painter O and Fink J~M
  2017 {\em Nature communications\/} {\bf 8} 1--7

\bibitem{Massel11}
Massel F, Heikkil{\"a} T, Pirkkalainen J~M, Cho S~U, Saloniemi H, Hakonen P~J
  and Sillanp{\"a}{\"a} M~A 2011 {\em Nature\/} {\bf 480} 351--354

\bibitem{Weis10}
Weis S, Rivi{\`e}re R, Del{\'e}glise S, Gavartin E, Arcizet O, Schliesser A and
  Kippenberg T~J 2010 {\em Science\/} {\bf 330} 1520--1523

\bibitem{Safavi11}
Safavi-Naeini A~H, Alegre T~M, Chan J, Eichenfield M, Winger M, Lin Q, Hill
  J~T, Chang D~E and Painter O 2011 {\em Nature\/} {\bf 472} 69--73

\bibitem{Teufel11a}
Teufel J~D, Li D, Allman M, Cicak K, Sirois A, Whittaker J and Simmonds R 2011
  {\em Nature\/} {\bf 471} 204--208

\bibitem{Hocke12}
Hocke F, Zhou X, Schliesser A, Kippenberg T~J, Huebl H and Gross R 2012 {\em
  New Journal of Physics\/} {\bf 14} 123037

\bibitem{Zhou13}
Zhou X, Hocke F, Schliesser A, Marx A, Huebl H, Gross R and Kippenberg T~J 2013
  {\em Nature Physics\/} {\bf 9} 179--184

\bibitem{Ockeloen16}
Ockeloen-Korppi C, Damsk{\"a}gg E, Pirkkalainen J~M, Heikkil{\"a} T, Massel F
  and Sillanp{\"a}{\"a} M 2016 {\em Physical Review X\/} {\bf 6} 041024

\bibitem{Bothner20}
Bothner D, Yanai S, Iniguez-Rabago A, Yuan M, Blanter Y~M and Steele G~A 2020
  {\em Nature communications\/} {\bf 11} 1--9

\bibitem{Cohen20}
Cohen M~A, Bothner D, Blanter Y~M and Steele G~A 2020 {\em Physical Review
  Applied\/} {\bf 13} 014028

\bibitem{Aspelmeyer14}
Aspelmeyer M, Kippenberg T~J and Marquardt F 2014 {\em Reviews of Modern
  Physics\/} {\bf 86} 1391

\bibitem{Agarwal10}
Agarwal G~S and Huang S 2010 {\em Physical Review A\/} {\bf 81} 041803

\bibitem{Gardiner04}
Gardiner C and Zoller P 2004 {\em Quantum noise: a handbook of Markovian and
  non-Markovian quantum stochastic methods with applications to quantum
  optics\/} (Springer Science \& Business Media)

\bibitem{Clerk10}
Clerk A~A, Devoret M~H, Girvin S~M, Marquardt F and Schoelkopf R~J 2010 {\em
  Reviews of Modern Physics\/} {\bf 82} 1155

\bibitem{Zhou19}
Zhou X, Cattiaux D, Gazizulin R, Luck A, Maillet O, Crozes T, Motte J~F,
  Bourgeois O, Fefferman A and Collin E 2019 {\em Physical Review Applied\/}
  {\bf 12} 044066

\bibitem{OConnell08}
O’Connell A~D, Ansmann M, Bialczak R~C, Hofheinz M, Katz N, Lucero E,
  McKenney C, Neeley M, Wang H, Weig E~M {\em et~al.\/} 2008 {\em Applied
  Physics Letters\/} {\bf 92} 112903

\end{thebibliography}
\end{document}